\documentclass[fleqn,twoside]{article}
\usepackage{espcrc1}

\newcommand{\AmS}{{\protect\the\textfont2
  A\kern-.1667em\lower.5ex\hbox{M}\kern-.125emS}}

\title{On combinatorial optimization for dominating sets
 (literature survey, new models)\\
  Preprint, Sep. 4, 2020\\
  DOI: 10.13140/RG.2.2.34919.68006}

\author{Mark Sh. Levin}

\begin{document}

\maketitle

\begin{abstract}
 The paper focuses on some versions of connected dominating set
 problems: basic problems and multicriteria problems.
 A literature survey on basic problem formulations
 and solving approaches is presented.
 The basic connected dominating set problems
 are illustrated by simplified numerical examples.
 New integer programming formulations
 of dominating set problems
  (with multiset estimates) are suggested.

~~

{\it Keywords:}~
 combinatorial optimization,
 connected dominating sets,
 multicriteria decision making,
 solving strategy,
 heuristics,
 networking,
 multiset

 ~~

\vspace{1pc}
\end{abstract}

\maketitle


\newcounter{cms}
\setlength{\unitlength}{1mm}

\section{Introduction}

 In recent decades, the significance of the
  connected dominating set problem has been increased
 (e.g., \cite{dud13,gar79,guha98,hay98,stojm01,yuj13}).
 This problem consists in searching for
 the minimum sized connected dominating set for the initial graph
 (e.g., \cite{dud13,gar79,guha98,hay98}).
 Main applications of the problems of this kind are pointed out in Table 1.

\begin{center}
\begin{small}
 {\bf Table 1.}
 Basic application domains of dominating set problems  \\
\begin{tabular}{| c | l | l |}
\hline
 No.&Application domain&Source(s)\\

\hline

 1.& Communication networks (e.g., design of virtual
  backbone in mobile  networks,
    &
    \cite{bai20,cheng03,dud13,erc07,hej12,oliv06}\\

    &connected dominating set based index scheme in WSNs)
& \cite{rama12,shit17,wuy10,yuj13}\\

 2.&Topology design for power (electricity) networked systems,
  electric power
     &\cite{cheng18,dorf06,hay02,liao15}\\

  & monitoring (e.g., minimum cardinality of a power dominating set)&\\

 3.&Converter placement problem in optical networks&\cite{dud13}\\

 4.& Placement of  monitoring devices
 (e.g., surveillance cameras, fire alarms)
    &\cite{henn12}\\

 & as total domination in graphs&\\

 5.& Social networks (positive influence dominating sets)
   & \cite{dinh14,ling18,wang11}\\

 6.&Multi-document summarization in information retrieval&\cite{potl13}\\

\hline
\end{tabular}
\end{small}
\end{center}

 This paper describes  some basic dominating set problems and
 connected dominating set problems (a brief survey) and some
 new problem formulations (with  multiset estimates).
 Some relations of dominating set problems and some other
 well-known combinatorial optimization problems
 are depicted in Fig. 1.
 Fig. 2 illustrates the dominating set problem
  (communication network application).

\section{Preliminaries}

 Let graph
 \(G=(A,R)\) be a connected undirected graph,
  where
 \(A\) is a vertex/node set
 and
 \(R\) is an edge set.
 A vertex subset \(B \subseteq A\)
is a {\it dominating set} if every vertex not in
 \(B\) has a neighbor in \(B\).
 The vertices from \(B\) are called dominators.
%
%
 If the subgraph induced by \(B\) is connected, then
 \(B\) is called a {\it connected dominating set}.
 Two special indices are considered:
 (a) {\it domination number}: \( \gamma(B) \)
 (minimum cardinality of \(B\)); and
 (b) {\it connected domination number}: \( \gamma_{c}(B) \)
 (minimum cardinality of connected \(B\)).
 Clearly, \( \gamma_{c}(B) \geq  \gamma (B)\).

 The basic dominating set (DS) problem is the following combinatorial optimization problem:

~~

 Find the minimum sized dominating set \(B\)
 (i.e., \( \min \gamma (B) \))
  for the initial graph \(G = (A,R)\).

~~

 The connected dominating set (CDS) problem is:

~~

 Find the minimum sized connected dominating set \(B\)
 (i.e., \( \min \gamma_{c} (B) \))
  for the initial graph \(G = (A,R)\).

~~

 The problems are proved to be NP-hard (e.g., \cite{dud13,gar79,hay98}).

\begin{center}
\begin{picture}(120,77)
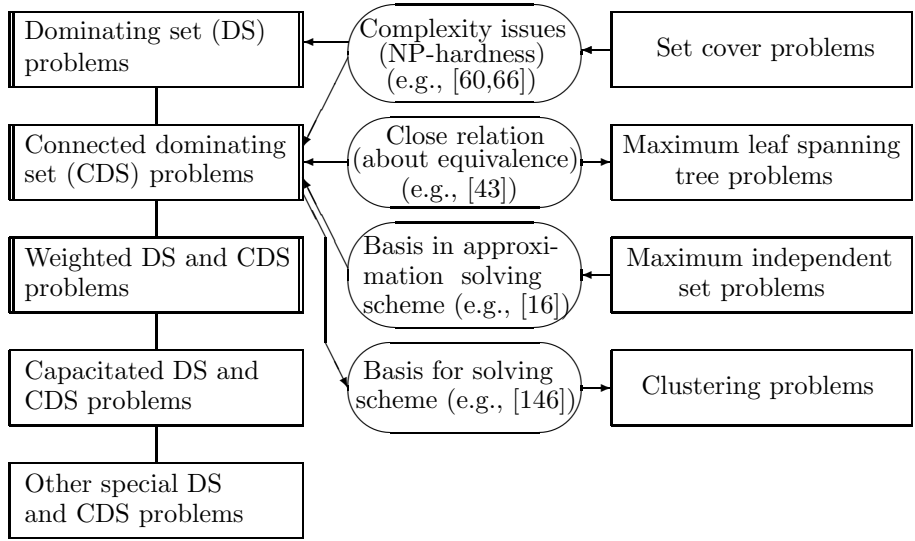
\put(01.2,00){\makebox(0,0)[bl]{Fig. 1. Some relations of
 dominating sets  and other combinatorial problems}}

\put(00,65){\line(1,0){39}} \put(00,75){\line(1,0){39}}
\put(00,65){\line(0,1){10}} \put(39,65){\line(0,1){10}}
\put(0.5,65){\line(0,1){10}} \put(38.5,65){\line(0,1){10}}

\put(02,70.5){\makebox(0,0)[bl]{Dominating set (DS)}}
\put(02,66.5){\makebox(0,0)[bl]{problems}}

\put(19.5,65){\line(0,-1){05}}
\put(00,50){\line(1,0){39}} \put(00,60){\line(1,0){39}}
\put(00,50){\line(0,1){10}} \put(39,50){\line(0,1){10}}
\put(0.5,50){\line(0,1){10}} \put(38.5,50){\line(0,1){10}}

\put(02,55.5){\makebox(0,0)[bl]{Connected dominating}}
\put(02,51.5){\makebox(0,0)[bl]{set (CDS) problems}}

\put(19.5,50){\line(0,-1){05}}
\put(00,35){\line(1,0){39}} \put(00,45){\line(1,0){39}}
\put(00,35){\line(0,1){10}} \put(39,35){\line(0,1){10}}
\put(0.5,35){\line(0,1){10}} \put(38.5,35){\line(0,1){10}}

\put(02,40.5){\makebox(0,0)[bl]{Weighted DS and CDS}}
\put(02,36.5){\makebox(0,0)[bl]{problems}}

\put(19.5,35){\line(0,-1){05}}
\put(00,20){\line(1,0){39}} \put(00,30){\line(1,0){39}}
\put(00,20){\line(0,1){10}} \put(39,20){\line(0,1){10}}

\put(02,25.5){\makebox(0,0)[bl]{Capacitated DS and}}
\put(02,21.5){\makebox(0,0)[bl]{CDS problems}}

\put(19.5,20){\line(0,-1){05}}
\put(00,05){\line(1,0){39}} \put(00,15){\line(1,0){39}}
\put(00,05){\line(0,1){10}} \put(39,05){\line(0,1){10}}

\put(02,10.5){\makebox(0,0)[bl]{Other special DS}}
\put(02,06.5){\makebox(0,0)[bl]{and CDS problems}}

\put(60.5,69.5){\oval(31,13)}

\put(46.5,71){\makebox(0,0)[bl]{Complexity issues}}
\put(49,67.5){\makebox(0,0)[bl]{(NP-hardness) }}
\put(50,64){\makebox(0,0)[bl]{(e.g., \cite{gar79,hay98})}}

\put(45,71){\vector(-1,0){06}}

\put(45,69){\vector(-1,-2){06}} \put(80,70){\vector(-1,0){04}}

\put(60.5,55){\oval(31,12)}

\put(50,57.5){\makebox(0,0)[bl]{Close relation}}
\put(45.5,53.5){\makebox(0,0)[bl]{(about equivalence)  }}

\put(52,49.5){\makebox(0,0)[bl]{(e.g., \cite{dud13}) }}

\put(45,55){\vector(-1,0){06}} \put(76,55){\vector(1,0){04}}

\put(60.5,40){\oval(31,14)}

\put(45,41){\vector(-1,2){06}}

\put(80,40){\vector(-1,0){04}}

\put(47,42.1){\makebox(0,0)[bl]{Basis in approxi-}}
\put(47,38.1){\makebox(0,0)[bl]{mation ~solving}}
\put(47,34.1){\makebox(0,0)[bl]{scheme  (e.g., \cite{bai20}) }}

\put(80,70){\vector(-1,0){04}}

\put(60.5,25){\oval(31,12)}

\put(39,51){\line(1,-2){03}} \put(42,45){\line(0,-1){14}}
\put(42,31){\vector(1,-2){03}}

\put(76,25){\vector(1,0){04}}

\put(47,25.5){\makebox(0,0)[bl]{Basis for solving}}
\put(46,21.5){\makebox(0,0)[bl]{scheme (e.g., \cite{yuj05})}}

\put(80,65){\line(1,0){40}} \put(80,75){\line(1,0){40}}
\put(80,65){\line(0,1){10}} \put(120,65){\line(0,1){10}}

\put(86,68.5){\makebox(0,0)[bl]{Set cover problems}}

\put(80,50){\line(1,0){40}} \put(80,60){\line(1,0){40}}
\put(80,50){\line(0,1){10}} \put(120,50){\line(0,1){10}}

\put(81.5,55.5){\makebox(0,0)[bl]{Maximum leaf spanning}}
\put(88.5,51.5){\makebox(0,0)[bl]{tree problems}}

\put(80,35){\line(1,0){40}} \put(80,45){\line(1,0){40}}
\put(80,35){\line(0,1){10}} \put(120,35){\line(0,1){10}}

\put(82,40.5){\makebox(0,0)[bl]{Maximum independent}}
\put(89,36.5){\makebox(0,0)[bl]{set problems}}

\put(80,20){\line(1,0){40}} \put(80,30){\line(1,0){40}}
\put(80,20){\line(0,1){10}} \put(120,20){\line(0,1){10}}

\put(84,23.5){\makebox(0,0)[bl]{Clustering problems}}

\end{picture}
\end{center}

\begin{center}
\begin{picture}(80,44.5)
\put(02,00){\makebox(0,0)[bl]{Fig. 2. Illustration for dominating
 set problem}}

\put(01,40.5){\circle*{1.4}} \put(01,40.5){\circle*{2.0}}
\put(02.8,39.5){\makebox(0,0)[bl]{Dominator}}

\put(01,35.5){\circle*{1.0}}

\put(02.8,35.5){\makebox(0,0)[bl]{Dominated}}
\put(02.8,32.5){\makebox(0,0)[bl]{element}}

\put(10,20){\oval(20,22)}

\put(04,27){\makebox(0,0)[bl]{Initial}}
\put(04,23){\makebox(0,0)[bl]{network}}
\put(04,19){\makebox(0,0)[bl]{(set of }}
\put(04,15){\makebox(0,0)[bl]{vertices/}}
\put(04,11){\makebox(0,0)[bl]{nodes)}}

\put(21,28){\vector(1,0){05}} \put(21,20){\vector(1,0){05}}
\put(21,12){\vector(1,0){05}}

\put(34.5,39){\makebox(0,0)[bl]{Resultant two-layer}}
\put(44,36){\makebox(0,0)[bl]{network}}

\put(50,29.5){\oval(26,11)}

\put(45,29){\circle*{1.4}} \put(45,29){\circle*{2.0}}
\put(50,26.5){\circle*{1.4}} \put(50,26.5){\circle*{2.0}}
\put(55,29){\circle*{1.4}} \put(55,29){\circle*{2.0}}

\put(45,29){\line(1,0){10}} \put(45,29.1){\line(1,0){10}}
\put(45,28.9){\line(1,0){10}}

\put(50,26.5){\line(-2,1){05}} \put(50,26.6){\line(-2,1){05}}
\put(50,26.4){\line(-2,1){05}}

\put(50,26.5){\line(2,1){05}} \put(50,26.6){\line(2,1){05}}
\put(50,26.4){\line(2,1){05}}

\put(38,30){\makebox(0,0)[bl]{Dominating set}}

\put(64,34){\makebox(0,0)[bl]{Network}}
\put(64,30){\makebox(0,0)[bl]{backbone, }}
\put(64,27){\makebox(0,0)[bl]{virtual}}
\put(64,23.5){\makebox(0,0)[bl]{backbone}}

\put(50,26.5){\line(-1,-3){4}} \put(50,26.5){\line(0,-1){12}}
\put(50,26.5){\line(1,-3){4}}

\put(50,14.5){\circle*{1}} \put(46,14.5){\circle*{1}}
\put(54,14.5){\circle*{1}}

\put(45,29){\line(-1,-3){5}} \put(45,29){\line(-1,-1){15}}
\put(45,29){\line(-2,-3){10}}

\put(30,14){\circle*{1}} \put(40,14){\circle*{1}}
\put(35,14){\circle*{1}}

\put(55,29){\line(1,-3){5}} \put(55,29){\line(1,-1){15}}
\put(55,29){\line(2,-3){10}}

\put(60,14){\circle*{1}} \put(65,14){\circle*{1}}
\put(70,14){\circle*{1}}

\put(50,11.5){\oval(45,12)}

\put(29.5,10){\makebox(0,0)[bl]{Elements under domination}}
\put(29.5,06.6){\makebox(0,0)[bl]{(e.g., terminals, end-users)}}

\end{picture}
\end{center}
  Evidently,
 the following basic requirements are examined for the problem
 (e.g., \cite{dud13,gar79,guha98,hay98,stojm01,yuj13}):
 (a) minimization of dominator set cardinality,
 (b) connectivity of the dominating vertices,
 (c) some special properties for the connectivity and domination
 (e.g., as \(k\)-connectivity, \(m\)-domination).

%
 Some simplified illustrative examples of \(k\)-connected dominating networks
 are shown in figures:

  (a)
   \(3\)-connected domination set example
    (\(3\)-connected dominating set is a clique, Fig. 3);

  (b) \(1\)-connected dominating set example
 (\(1\)-connected dominating set is a tree, Fig. 4);

 (c) \(2\)-connected dominating set example
 (\(2\)-connected dominating set is a ring, Fig. 5);

  (d) \(2\)-connected \(3\)-dominating example
 (i.e, (\(2,3\))-CDS problem,
  \(2\)-connected dominating set is a ring, Fig. 6).

 In addition,
 some special requirements are considered as well (Table 2).
 Note, the requirements can be used as constraints or as criteria
 in multicriteria (multiobjective) problem formulations.

~~

~~

\begin{center}
\begin{picture}(77,47)
\put(00,00){\makebox(0,0)[bl]{Fig. 3.
 \(3\)-connected domination set example}}

\put(15,35){\line(1,0){20}} \put(15,20){\line(1,0){20}}
\put(15,35.2){\line(1,0){20}} \put(15,20.2){\line(1,0){20}}
\put(15,34.8){\line(1,0){20}} \put(15,19.8){\line(1,0){20}}

\put(15,20){\line(0,1){15}} \put(35,20){\line(0,1){15}}
\put(15.2,20){\line(0,1){15}} \put(35.2,20){\line(0,1){15}}
\put(14.8,20){\line(0,1){15}} \put(34.8,20){\line(0,1){15}}

\put(15,20){\line(4,3){20}} \put(35,20){\line(-4,3){20}}
\put(15,20.2){\line(4,3){20}} \put(35,20.2){\line(-4,3){20}}
\put(15,19.8){\line(4,3){20}} \put(35,19.8){\line(-4,3){20}}

\put(09,32.5){\oval(18,18)}
\put(02,42.5){\makebox(0,0)[bl]{Cluster 1}}

\put(15,35){\circle*{2}}

\put(15,35){\line(-2,1){10}} \put(15,35){\line(-1,-1){10}}
\put(15,35){\line(-2,-1){10}} \put(15,35){\line(-1,0){10}}

\put(05,40){\circle*{1}} \put(05,35){\circle*{1}}
\put(05,30){\circle*{1}} \put(05,25){\circle*{1}}

\put(39.5,35){\oval(16,16)}

\put(48,36.5){\makebox(0,0)[bl]{Cluster}}
\put(52.5,32.5){\makebox(0,0)[bl]{2}}

\put(15,35){\circle*{2}} \put(35,35){\circle*{2}}

\put(35,35){\line(2,1){10}} \put(35,35){\line(0,1){05}}
\put(35,35){\line(2,-1){10}} \put(35,35){\line(1,0){10}}

\put(45,35){\circle*{1}} \put(45,40){\circle*{1}}
\put(45,30){\circle*{1}} \put(35,40){\circle*{1}}

\put(12.5,15){\oval(23,14)}

\put(01,05){\makebox(0,0)[bl]{Cluster 3}}

\put(15,20){\circle*{2}}

\put(15,20){\line(-1,0){10}} \put(15,20){\line(-1,-1){10}}
\put(15,20){\line(0,-1){10}} \put(15,20){\line(1,-2){05}}

\put(05,20){\circle*{1}} \put(05,10){\circle*{1}}
\put(15,10){\circle*{1}} \put(20,10){\circle*{1}}

\put(41.5,17.5){\oval(16,18)}

\put(50,19.5){\makebox(0,0)[bl]{Cluster}}
\put(54.5,15.5){\makebox(0,0)[bl]{4}}

\put(35,20){\circle*{2}}

\put(35,20){\line(2,-1){10}} \put(35,20){\line(1,0){10}}
\put(35,20){\line(2,1){10}} \put(35,20){\line(1,-2){05}}

\put(45,15){\circle*{1}} \put(45,20){\circle*{1}}
\put(45,25){\circle*{1}} \put(40,10){\circle*{1}}

\end{picture}
%
\begin{picture}(68,42)
\put(00,00){\makebox(0,0)[bl]{Fig. 4.
 \(1\)-connected domination set example}}

\put(15,25){\circle*{2}}

\put(15,25){\line(-2,1){10}} \put(15,25){\line(-1,-1){10}}
\put(15,25){\line(-2,-1){10}} \put(15,25){\line(-1,0){10}}

\put(05,30){\circle*{1}} \put(05,25){\circle*{1}}
\put(05,20){\circle*{1}} \put(05,15){\circle*{1}}

\put(14.9,25){\line(-1,-3){05}} \put(14.7,25){\line(-1,-3){05}}
\put(14.5,25){\line(-1,-3){05}}

\put(15.1,25){\line(2,-3){10}} \put(15.3,25){\line(2,-3){10}}
\put(15.5,25){\line(2,-3){10}}

\put(30,35){\circle*{2}}

\put(30,35){\line(0,1){05}} \put(30,35){\line(1,1){05}}
\put(30,35){\line(-1,1){05}} \put(30,35){\line(2,1){10}}

\put(30,40){\circle*{1}} \put(35,40){\circle*{1}}
\put(25,40){\circle*{1}} \put(40,40){\circle*{1}}

\put(29.9,35){\line(-3,-2){15}} \put(29.7,35){\line(-3,-2){15}}
\put(29.5,35){\line(-3,-2){15}}

\put(30.1,35){\line(3,-2){15}} \put(30.3,35){\line(3,-2){15}}
\put(30.5,35){\line(3,-2){15}}

\put(45,25){\circle*{2}}

\put(45,25){\line(2,1){10}} \put(45,25){\line(1,1){05}}
\put(45,25){\line(2,-1){10}} \put(45,25){\line(1,0){10}}

\put(55,25){\circle*{1}} \put(55,30){\circle*{1}}
\put(55,20){\circle*{1}} \put(50,30){\circle*{1}}

\put(44.7,25){\line(-1,-3){05}} \put(44.9,25){\line(-1,-3){05}}
\put(44.5,25){\line(-1,-3){05}}

\put(45.3,25){\line(2,-3){10}} \put(45.5,25){\line(2,-3){10}}
\put(45.1,25){\line(2,-3){10}}

\put(10,10){\circle*{2}}

\put(10,10){\line(-1,0){10}} \put(10,10){\line(-1,-1){05}}
\put(10,10){\line(0,-1){05}} \put(10,10){\line(1,-1){05}}

\put(00,10){\circle*{1}} \put(05,05){\circle*{1}}
\put(10,05){\circle*{1}} \put(15,05){\circle*{1}}

\put(55,10){\circle*{2}}

\put(55,10){\line(2,-1){10}} \put(55,10){\line(1,0){10}}
\put(55,10){\line(2,1){10}} \put(55,10){\line(1,-1){05}}

\put(65,05){\circle*{1}} \put(65,10){\circle*{1}}
\put(65,15){\circle*{1}} \put(60,05){\circle*{1}}

\put(25,10){\circle*{2}}

\put(25,10){\line(-1,-1){05}} \put(25,10){\line(0,-1){05}}
\put(25,10){\line(-1,-2){02.5}} \put(25,10){\line(1,-1){05}}

\put(25,05){\circle*{1}} \put(30,05){\circle*{1}}
\put(22.5,05){\circle*{1}} \put(20,05){\circle*{1}}

\put(40,10){\circle*{2}}

\put(40,10){\line(-1,-1){05}} \put(40,10){\line(0,-1){05}}
\put(40,10){\line(-1,-2){02.5}} \put(40,10){\line(1,-1){05}}

\put(40,05){\circle*{1}} \put(35,05){\circle*{1}}
\put(37.5,05){\circle*{1}} \put(45,05){\circle*{1}}

\end{picture}
\end{center}

\begin{center}
\begin{picture}(77,32)
\put(00,00){\makebox(0,0)[bl]{Fig. 5.
 \(2\)-connected domination set example}}

\put(15,25){\line(1,0){35}} 
\put(15,25.2){\line(1,0){35}} 
\put(15,24.8){\line(1,0){35}} 

\put(15,10){\line(1,0){35}} \put(15,10.2){\line(1,0){35}}
\put(15,09.8){\line(1,0){35}}

\put(15,10){\line(0,1){15}} \put(50,10){\line(0,1){15}}
\put(15.2,10){\line(0,1){15}} \put(50.2,10){\line(0,1){15}}
\put(14.8,10){\line(0,1){15}} \put(49.8,10){\line(0,1){15}}

\put(15,25){\circle*{2}}

\put(15,25){\line(-2,1){10}} \put(15,25){\line(-1,-1){10}}
\put(15,25){\line(-2,-1){10}} \put(15,25){\line(-1,0){10}}

\put(05,30){\circle*{1}} \put(05,25){\circle*{1}}
\put(05,20){\circle*{1}} \put(05,15){\circle*{1}}

\put(50,25){\circle*{2}}

\put(50,25){\line(2,1){10}} \put(50,25){\line(0,1){05}}
\put(50,25){\line(2,-1){10}} \put(50,25){\line(1,0){10}}

\put(60,25){\circle*{1}} \put(60,30){\circle*{1}}
\put(60,20){\circle*{1}} \put(50,30){\circle*{1}}

\put(15,10){\circle*{2}}

\put(15,10){\circle*{2}}

\put(15,10){\line(-1,0){10}} \put(15,10){\line(-1,-1){05}}
\put(15,10){\line(0,-1){05}} \put(15,10){\line(1,-2){02.5}}

\put(15,05){\circle*{1}} \put(10,05){\circle*{1}}
\put(05,10){\circle*{1}} \put(17.5,05){\circle*{1}}

\put(50,10){\circle*{2}}

\put(50,10){\line(2,-1){10}} \put(50,10){\line(1,0){10}}
\put(50,10){\line(2,1){10}} \put(50,10){\line(1,-1){05}}

\put(60,05){\circle*{1}} \put(60,10){\circle*{1}}
\put(60,15){\circle*{1}} \put(55,05){\circle*{1}}

\put(32.5,10){\circle*{2}}

\put(32.5,10){\line(0,-1){05}} \put(32.5,10){\line(-1,-1){05}}
\put(32.5,10){\line(1,-1){05}} \put(32.5,10){\line(-1,-2){02.5}}

\put(32.5,05){\circle*{1}} \put(37.5,05){\circle*{1}}
\put(30,05){\circle*{1}} \put(27.5,05){\circle*{1}}

\put(32.5,25){\circle*{2}}

\put(32.5,25){\line(0,1){05}} \put(32.5,25){\line(-1,2){02.5}}
\put(32.5,25){\line(-1,1){05}} \put(32.5,25){\line(1,1){05}}

\put(32.5,30){\circle*{1}} \put(37.5,30){\circle*{1}}
\put(30,30){\circle*{1}} \put(27.5,30){\circle*{1}}

\end{picture}
%
\begin{picture}(66,32)
\put(00,00){\makebox(0,0)[bl]{Fig. 6.
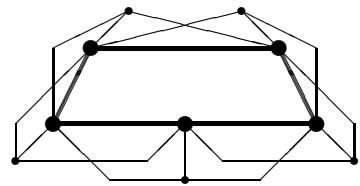
 \(2\)-connected \(3\)-domination set example}}

\put(20,25){\line(1,0){25}} \put(20,25.2){\line(1,0){25}}
\put(20,24.8){\line(1,0){25}}

\put(15,15){\line(1,0){35}} \put(15,15.2){\line(1,0){35}}
\put(15,14.8){\line(1,0){35}}

\put(15,15){\line(1,2){05}} \put(50,15){\line(-1,2){05}}
\put(15.2,15){\line(1,2){05}} \put(50.2,15){\line(-1,2){05}}
\put(14.8,15){\line(1,2){05}} \put(49.8,15){\line(-1,2){05}}

\put(20,25){\circle*{2}}

\put(25,30){\circle*{1}}

\put(25,30){\line(-1,-1){05}} \put(25,30){\line(4,-1){20}}
\put(25,30){\line(-2,-1){10}} \put(15,25){\line(0,-1){10}}

\put(45,25){\circle*{2}}

\put(40,30){\circle*{1}}

\put(40,30){\line(1,-1){05}} \put(40,30){\line(-4,-1){20}}
\put(40,30){\line(2,-1){10}} \put(50,25){\line(0,-1){10}}

\put(15,15){\circle*{2}}

\put(10,10){\circle*{1}}

\put(10,10){\line(1,1){05}}

\put(10,10){\line(0,1){05}} \put(10,15){\line(1,1){10}}
\put(10,10){\line(1,0){17.5}} \put(27.5,10){\line(1,1){05}}

\put(50,15){\circle*{2}}

\put(55,10){\circle*{1}}

\put(55,10){\line(-1,1){05}}

\put(55,10){\line(0,1){05}} \put(55,15){\line(-1,1){10}}
\put(55,10){\line(-1,0){17.5}} \put(37.5,10){\line(-1,1){05}}

\put(32.5,15){\circle*{2}}

\put(32.5,07.5){\circle*{1}} \put(32.5,7.5){\line(0,1){07.5}}

\put(32.5,07.5){\line(-1,0){10}} \put(22.5,07.5){\line(-1,1){7.5}}
\put(32.5,07.5){\line(1,0){10}} \put(42.5,07.5){\line(1,1){7.5}}


\end{picture}
\end{center}

\begin{center}
\begin{small}
 {\bf Table 2.} Additional special requirements in dominating set problems\\
\begin{tabular}{| c | l | l |}
\hline
 No.&Requirement&Source(s)\\

\hline

 1.& Restricted diameter on connected dominating set
    &\cite{buch14}\\

 2.& Multipoint relays& \cite{adj05,rama12,wuj06}\\

 3.& Routing cost constraint in networks&\cite{duh10,duh12a,duh16}\\

 4. &Demand constraints &\cite{kao11}\\

 5. &Capacity constraints &\cite{kao11}\\

 6.&Fault-tolerance  of connected dominating set&\cite{zhouj14}\\

\hline
\end{tabular}
\end{small}
\end{center}

 Two combinatorial problems are very close to the examined problems:

~~

 {\bf Close problem 1.} The  maximum independent set problem
 (e.g., \cite{alv19,alz03,andr12,bai20,gar79,moha16,sunx19,wuw06}):

~~

  Find a maximum subset of vertices of an input graph
 (an independent or stable set)
 such that there is no edges between two vertices in the subset.

~~

 This problem is often used as a preliminary one in approximation
 two-phase approach for the designing the minimum connected dominating sets
 (e.g., \cite{bai20,funke06,min06,wanp08}):
   (1) to construct a maximal independent set for the initial network;
   (2) to connect the nodes in it.

~~

 {\bf Close problem 2.} The maximum leaves spanning tree problem:
 (e.g., \cite{alon09,caroy00,fuj03,gar79,kleit91}):

~~

 Find a spanning tree of an input graph so that the number of the
 tree leaves is maximal.

~~

 This problem is equivalent to the problem of computing
 \(\gamma_{c}(G)\),
 because a vertex subset is a {\it connected dominating set}
 if and only its compliment is (contained in) the set of leaves
 of a spanning tree (e.g., \cite{caroy00}).

~~~

 Generally, the above-mentioned two problems
 belong the class of NP-hard problems as well
 (e.g., \cite{gar79,hay98}).
 Thus, for the problems
 (i.e., minimum connecting dominating set, maximum independent set,
 maximum leaves spanning tree)
  the following approaches are used:
 (1) exact enumerative methods
 (e.g., Branch-and-Bound method)
  (e.g., \cite{dud13,fuj03,simo11});
%
(2) approximation heuristics
   (e.g., \cite{but04,guha98,ruan04,zou11});
 and
 (3) metaheuristics and hybrid methods
    (e.g., \cite{alb18,bai20,potl13}).
%
 For some special cases of the problems polynomial approaches are suggested:
 (a) polynomial algorithms
(e.g., \cite{henn20,lap13,panda13,prad19,qiao03}),
  and
 (b) polynomial time approximate schemes (PTAS)
 (e.g., \cite{cheng03,gaox10,kao11,nieb05,zhang09}).

 Some basic versions of connected set problems
 are listed in Table 3
 (e.g., \cite{dud13,gar79,guha98,hay98,stojm01,yuj13}).

\newpage
\begin{center}
\begin{small}
 {\bf Table 3.}
 Basic dominating set and connected dominating set problems, part 1  \\
\begin{tabular}{| c | l | l |}
\hline
 No.&Problem type&Source(s)\\

\hline
 1.&Basic  surveys on problems and applications:& \\

 1.1.&Connected dominating set: theory and applications&\cite{dud13}\\

 1.2.&Connected dominating sets in wireless ad hoc and sensor networks
     &\cite{yuj13}\\

 1.3.& Connected dominating set in sensor networks and MANETs
   &\cite{blum05}\\

 1.4.&Connected dominating set in wireless networks&\cite{duh13}\\

\hline
 2. &Basic problems:&\\

 2.1.&Dominating set problem, minimum dominating set problem
   &\cite{dud13,gar79,hoc06}\\

 2.2.&Independent dominating sets in  graphs,
 &\cite{godd12,irv91,liu15,rad13}\\

 &minimum independent dominating set&\\

 2.3.&Dominating sets in planar graphs&\cite{fom06}\\

 2.4.&Connected dominating set problems (e.g., minimum connected
 &\cite{bai20,blum05,buch14,dud13,duh13,gar79,guha98,linz12}\\

   &dominating set,
     i.e., minimum cardinality of the dominating set)
  &\cite{moha16,moha17,oliv06,ruan04,scha12,shit17}\\

  &&\cite{simo11,sunx19,wuj01,wuw06}\\

 2.5.&Connected dominating set problems in unit disk graphs
  &\cite{dud13,fuku18,shiy19,zou11} \\

 2.6.&Planar connected dominating set problem&\cite{lok11}\\

\hline
 3.&Dominating set problems with special kinds of connectivity&\\

  &(e.g., weakly, strongly, \(k\)-connected):&\\

 3.1.&Minimum size weakly-connected dominating sets
   &\cite{akba10a,alz03,cheny02,cheny05,dud13,yuj12}\\

 3.2.&Strongly connected dominating sets in
  networks with unidirectional links
   &\cite{dud06b,dud13}\\

 3.3.&Total domination set problems
    &\cite{allan84,changm97,cock80,henn12,henn13,henn13b,scha12}\\

 3.4.&Problems on total \(k\)-domination in graphs
  & \cite{berm19a,berm19b}\\

 3.5.&Problem on semitotal domination in graphs&\cite{henn14}\\

 3.6.&Problem on double domination in graphs&\cite{haj19,haran05,har00}\\

 3.7.&Mixed domination problem in graphs&\cite{lan13}\\

 3.8.&\(k\)-connected \(m\)-dominating set problem
  ((\(k,m\))-CDM problem)
   & \cite{dai06,fuku18,liy12,shangw08,thai07}\\

 3.9.&\(k\)-connected \(m\)-dominating set problem
  ((\(k,m\))-CDM problem)
   & \cite{fuku18,nutov17,shiy19}\\

 &with node weights&\\

\hline
 4.&Weighted dominating set problems:&\\

 4.1.&Node weighted connected dominating set problem (e.g., vertex
 &\cite{alb18,changm97,dud13,guha98,ling16,potl13}\\

  &importance) (minimum weight connected dominating set problem)
  &\cite{vaz18,wangy16,zou11}\\

 4.2.&Minimum weight \(k\)-connected \(m\)-fold dominating set
   &\cite{fuku18,nutov17,shiy19}\\

 &(minimum weight \((k,m)\)-CDS problem)&\\

 4.3.&Optimal degree constrained minimum-weight connected dominating
  &\cite{akba12}\\

 &set problem (network backbone formation)&\\

 4.4.&Weighted connected dominating set problem in unit disk graphs
  &\cite{dud13} \\

 4.5.&Minimum weight partial connected set cover problem
   & \cite{liang16}\\

\hline
 5.&Capacitated domination problems:&\\

 5.1.&Capacitated domination set problems&\cite{dom08,lied14}\\

 5.2.&Minimum capacitated dominating set problem&\cite{pina19}\\

 5.3.&Capacitated (soft) domination problem (minimum cardinality of DS
   &\cite{kao11}\\

 &satisfying both the capacity and demand constraints)&\\

 5.4.&Capacitated \(b\)-edge dominating set problem&\cite{berg07}\\

\hline
 6.&Edge dominating set problems:&\\

 6.1.&Edge dominating set problems
   &\cite{chle06,fuj01,gar79,hort93,yan80}\\

 6.2.&\(2\)-edge connected dominating sets of a graph&\cite{lih16}\\

 6.3.& \(b\)-edge dominating set problem&\cite{fuk05}\\

 6.4.&Capacitated \(b\)-edge dominating set problem&\cite{berg07}\\

 6.5.&Edge total domination problems&\cite{zhao14}\\

 6.6.&Edge weighted dominating set problem&\cite{carr00,fuj02}\\

 6.7.&Edge weighted connected dominating set problem&\cite{guha98}\\

\hline
 7.&Multi-hop dominating set problems:&\\

 7.1.&Hop domination in graphs (\(k\)-step dominating sets)
   &\cite{henn20b,kund16}\\

 7.2.&Connected \(k\)-hop dominating set problems
   &\cite{coel17,gaox10,ngu06,nocc03}\\

 & &\cite{wuy09,yangh08,zhengc11}\\

 7.3.&Connected dominating sets with multipoint relays
   &\cite{adj05,rama12,wuj06}\\

 7.4.& Energy-efficient dominating tree
 (in multi-hop wireless networks)
   & \cite{yur09}\\

 7.5.&Connected dominating set
 for multi-hop wireless networks
  &\cite{suren15}\\

\hline
\end{tabular}
\end{small}
\end{center}

\newpage
\begin{center}
\begin{small}
 {\bf Table 3.}
 Basic dominating set and connected dominating set problems, part 2  \\
\begin{tabular}{| c | l | l |}
\hline
 No.&Problem type&Source(s)\\

\hline
 8.&Steiner weighted dominating set problems:&\\

 8.1.&Steiner connected dominating set problem
 &\cite{tork15,guha98,min06,wuy05}\\

 8.2.&Node weighted Steiner connected  dominating set problem&\cite{torke11,guha98}\\

 &(minimum weight Steiner connected dominating set problem)&\\

\hline
 9.&Special dominating set problems:&\\

 9.1.&\(k\)-fair domination in graphs&\cite{caro12}\\

 9.2.&Energy efficient stable connected dominating set&\cite{rama12}\\

 9.3.&Minimum connected dominating set for certain circulant networks
  &\cite{parth15} \\

 9.4.&Routing-cost constrained connected dominating set problems
   &\cite{duh16}\\

 9.5.&Highly connected multi-dominating sets problems&\cite{fuku18}\\

 9.6.&Load balanced connected dominating set problem
   &\cite{hej12}\\

 9.7.&Dominating sets and connected dominating sets in dynamic graphs
   &\cite{hju19}\\

 9.8.&Reconfiguration of dominating sets&\cite{suz16}\\

\hline
\end{tabular}
\end{small}
\end{center}


 Table 4 contains a list of basic solving approaches.

\begin{center}
\begin{small}
 {\bf Table 4.} Basic solving approaches, part 1 \\
\begin{tabular}{| c | l | l |}
\hline
 No.  &Approach & Source(s) \\

\hline
 1.&Basic  surveys on problems and applications:& \\

 1.1.&Survey on problems and solving approaches &\cite{dud13}\\

 1.2.&Classification and comparison of connected DS construction algorithms
  & \cite{yuj13} \\

\hline
 2.&Exact methods:&\\

 2.1.&Branch-and-Bound algorithm for
    minimum connected dominating set problem
    &\cite{simo11}\\

 2.2.&Polynomial time algorithms for
  minimum paired-dominating sets (special graphs:
  &\cite{henn20,lap13,panda13}\\

  & tree, convex bipartite graph,
  strongly orderable graph, permutation graph)
   &\cite{prad19,qiao03}\\

 2.3.& Enumeration algorithms for minimal dominating sets problems
  &\cite{cout13}\\

 2.4.&  Exact algorithms for dominating set (exponential time algorithms,
 &\cite{fom04,lied14,roo11}\\

 & Branch and reduce algorithm)&\\

 2.5.&Exact algorithm for connected red-blue dominating set&\cite{abu11}\\

\hline
 3.&Heuristics, approximation algorithms:&\\

 3.1.&Heuristic for the minimum connected dominating
 set problem
   &\cite{but04}\\

 &on ad hoc wireless networks &\\

 3.2.&Extended localized algorithm for connected dominating
   set formation
  &\cite{daif04}\\

 &in Ad Hoc wireless networks &\\

 3.3.&Approximation algorithm for minimum size weakly-connected dominating sets
   &\cite{cheny02}\\

 3.4.&Efficient algorithms for the minimum connected
    domination on trapezoid graphs
  &\cite{tsai07}\\

 3.4.&Greedy approximation for minimum connected dominating sets
  &\cite{ruan04}\\

 3.5.&Approximations for minimum-weighted dominating sets&\cite{zou11}\\

    &and minimum-weighted connected dominating sets&\\

 3.6.&Approximating minimum size weakly-connected dominating sets
    &\cite{cheny02}\\

 3.7.&Approximation algorithms for
  \(1\)-\(m\)-CDS and \(k\)-\(k\)-CDM problems
    &\cite{thai07}\\

 3.8.&Approximation algorithms for
 highly connected multi-dominating sets problems
   &\cite{fuku18}\\

 3.9.&Constant-approximation for minimum-weight
 (connected) dominating sets
   & \cite{amb06}\\

 3.10.&Approximation for \(k\)-hop connected dominating set problem
   & \cite{coel17}\\

 3.11.&Two-phase approximation algorithms (for minimum CDS)
   & \cite{wanp08}\\

 3.12.&Performance guaranteed approximation algorithm for minimum
 \(k\)-connected
   & \cite{zhang16}\\

 & \(m\)-fold dominating set&\\

 3.13.&Approximation centralized algorithm for minimum CDS
  & \cite{luoc18}\\

 & constructing virtual backbones in WSNs&\\

 3.14.&Approximation distributed algorithm for minimum CDS
  (in unit disk graphs)
  &\cite{funke06}\\

 3.15.&Simple heuristic for minimum connected dominating
 set in graphs
   &\cite{wanp03} \\

 & (approximation algorithm)&\\

 3.16.&Approximation algorithm as distributed construction
 of connected dominating set
   & \cite{jal17}\\

  &(unit disk graphs)&\\

 3.17.&Efficient randomized distributed greedy algorithm for constructing
 small
 &\cite{jia02}\\

 &dominating sets (approximation)&\\

 3.18.&Approximation algorithms for \(k\)-connected
 \(m\)-dominating set problems
   & \cite{nutov17}\\

\hline
\end{tabular}
\end{small}
\end{center}

\newpage
\begin{center}
\begin{small}
 {\bf Table 4.} Basic solving approaches, part 2 \\
\begin{tabular}{| c | l | l |}
\hline
 No.  &Approach & Source(s) \\

\hline
 4.&Metaheuristics:&\\

 4.1.&Efficient mathheuristic for the minimum-weight
 dominating set problem.
   &\cite{alb18}\\

 4.2.&Hybrid metaheuristic algorithms for minimum weight dominating set
    &\cite{potl13}\\

 4.3.&Enhanced ant colony optimization metaheuristic
 (minimum dominating set problem)
 &\cite{hoc06}\\

 4.4.&Two stage efficient centralized algorithm for connected
 dominating set
 &\cite{fud15}\\

 &on wireless networks&\\

\hline
 5.&Evolutionary methods:&\\

 5.1.&Hybrid genetic algorithm for minimum dominating set problem
   &\cite{hed10}\\

 5.2.&Hybrid GA for minimum weight connected dominating set problem
  &\cite{dag17}\\

 5.3.&Effective hybrid memetic algorithm for the minimum
  weight dominating set problem
   &\cite{ling16}\\

 5.4.&Memetic algorithm for finding
 positive influence dominating sets in social networks
  &\cite{ling18}\\

\hline
 6.&Linear time algorithms:&\\

 6.1.&Linear time algorithm for optimal
 \(k\)-hop dominating set problem
  &\cite{henn20b,kund16}\\

 & (for tree, for bipartite permutation graphs)&\\

 6.2.&Linear time algorithms for finding a dominating set
 of fixed size in degenerated graphs
  &\cite{alon09b}\\

 6.3.& Linear-time exact algorithm for tree
 (positive influence dominating set in social networks)
  &\cite{dinh14}\\

 6.4.&Linear time algorithms for generalized edge
 dominating set problems
  &\cite{berg08}\\

 6.5.&Dynamic programming style  linear time algorithm for
 \(k\)-power domination
 &\cite{cheng18}\\

 &problem in weighted trees&\\

\hline
 7.&Polynomial time approximation schemes:&\\

 7.1.&PTAS for minimum
   connected dominating set in ad hoc wireless networks
   & \cite{cheng03}\\

 7.2.&PTAS for minimum connected dominating set
  in \(3\)-dimensional WSNs
  &\cite{zhang09}\\

 7.3.&PTAS for minimum \(d\)-hop connected dominating set
 in growth-bounded graphs
 &  \cite{gaox10}\\

 7.4.&PTAS for the minimum dominating set problem
 in unit disk graphs
 & \cite{nieb05}\\

 7.5.&PTAS for minimum connected dominating set with routing
 cost constraint in WSNs
    & \cite{duh10,duh12a}\\

 7.6.&PTAS for routing-cost constrained minimum connected
 dominating set
  &\cite{wul15}\\

 &(in growth bounded graphs)&\\

 7.7.&PTAS for capacitated domination problem on trees&\cite{kao11}\\

 7.8.&PTAS for the minimum weighted dominating set in
 growth bounded graphs
   &\cite{wangz12}\\

\hline
 8.&Special approaches:&\\

 8.1.&Search tree technique for dominating set on planar graphs
   & \cite{alber05}\\

 8.2.&Distributed algorithm for efficient construction and
 maintenance
  &\cite{yangh08}\\

   &of connected \(k\)-hop dominating set&\\

 8.3.&Distributed learning automata approach for
  weakly connected dominating set
  &\cite{akba10a}\\

 8.4.& Adaptive algorithms for connected dominating sets
 & \cite{fuku19a}\\

\hline
\end{tabular}
\end{small}
\end{center}

\section{Integer programming formulations of dominating set problems}

 The considered dominating set problem formulations
 are listed in Table 5.

\subsection{Basic models}

  The minimum dominating set problem consists in
   finding the minimum dominating set \(B \subseteq A\)
  in graph \(G = (A,R)\), where
 \(A = \{a_{1},...,a_{i},...,a_{n}\}\),
 \(R = \{r_{1},...,r_{j},...,r_{q}\}\).
  The following binary variable \(x_{i}\)
  (\(i=\overline{1,n}\))  is used:
  \( \forall a_{i} \in A\)
  \(x_{i} = 1\) if \(a\) is selected for \(B\)
 (i.e., \(a \in B\)) and
 \(x_{i} = 0\) otherwise.
 Thus, the solution of the dominating set problems is defined by
 binary vector
 \(\overline{x} = (x_{1},...,x_{i},...,x_{n})\).
 The following optimization models can be considered:

~~

 {\bf Model 1.} The integer programming formulation of the basic minimum dominating set problem is:
 \[ \min \gamma (b) = \min \sum_{i=1}^{n} ~x_{i}\]
 \[s.t.~~~~ \forall a_{\zeta_{1}} \in (A \backslash B) ~~
 ~~\exists a_{\zeta_{2}} \in  B~ (i.e., ~ x_{a_{\zeta_{2}}} = 1)
 ~~such~~that ~~ (a_{\zeta_{1}},a_{\zeta_{2}}) \in R .\]

~~~

 {\bf Model 2.} The integer programming formulation of the basic
 minimum connected dominating set problem is
 (here \(B_{c} \subseteq A \) is the connected dominating set):
 \[ \min \gamma_{c} (b) = \min \sum_{i=1}^{n} ~x_{i}\]
 \[s.t.~~~~ \forall a_{\zeta'}, a_{\zeta''} \in B_{c}
 ~~\exists ~~ path ~~ l(a_{\zeta'}, a_{\zeta''})=<a_{\zeta'},...,a_{\zeta''}> , \]
 \[\forall a_{\zeta_{1}} \in (A \backslash B_{c}) ~~
 ~~\exists a_{\zeta_{2}} \in  B_{c}~ (i.e., ~ x_{a_{\zeta_{2}}} = 1)
 ~~such~~that ~~ (a_{\zeta_{1}},a_{\zeta_{2}}) \in R .\]

~~~

 In addition,
 vertex weights are considered:
 \(w(a_{i}) > 0\)  (\(\forall a_{i} \in A\)).

~~

\begin{center}
\begin{small}
 {\bf Table 5.}
 Integer programming formulations of dominating set problems
 \\
\begin{tabular}{| c | l | }
\hline
 No. & Model   \\

\hline

 I.&Basic models:\\

 Model 1&Integer programming formulation of the basic minimum dominating set problem\\

 Model 2&Integer programming formulation of the basic
 minimum connected dominating set problem\\

 Model 3&
  Integer programming formulation of the minimum node weighted dominating set
  problem\\

 Model 4&
 Integer programming formulation of the
 minimum vertex weighted connected\\

  &dominating set problem\\

\hline
 II.&Basic multicriteria models:\\

 Model 5&
 Multicritiria (multiobjective)
 minimum node weighted dominating set problem\\

 Model 6&
 Multicriteria (multiobjective) minimum node weighted connected dominating set
 problem\\

\hline
 III.&Models of \(k\)-connected dominating set problems:\\

 Model 7&
 Integer programming formulation of the basic
 minimum  \(k\) connected dominating set problem\\

 Model 8&
 Integer programming formulation of the
 minimum vertex weighted \(k\)-connected connected\\

 &dominating set problem\\

 Model 9&Multicriteria (multiobjective) minimum node weighted
 \(k\)-connected dominating set problem\\

\hline
 IV.&Models of \(k\)-connected \(m\)-dominating set problems:\\

 Model 10&
 Integer programming formulation of the basic
 minimum  \(k\)-connected \(m\)-dominating set problem\\

 Model 11&
 Integer programming formulation of the
 minimum vertex weighted \(k\)-connected connected\\

 & \(m\)-dominating set problem\\

 Model 12&
 Multicriteria (multiobjective) minimum node weighted \(k\)-connected
  \(m\)-dominating set problem\\

\hline
 V.&Dominating problems with multiset estimates:\\

 Model 13&
 Minimum node weighted dominating set problem with multiset
 estimates\\

 & (the objective function is based on ``generalized median'')\\

 Model 14&
 Minimum node weighted connected dominating set problem with multiset
 estimates\\

 &(the objective function is based on ``generalized median'')\\

 Model 15&
 Multicriteria (multiobjective) minimum node weighted \(k\)-connected
  dominating set problem\\

 & with multiset estimates
  (the objective function is based on ``generalized median'')\\

 Model 16&
 Multicriteria (multiobjective) minimum node weighted \(k\)-connected
  \(m\)-dominating set problem \\

 & with multiset estimates
  (the objective function is based on ``generalized median'')\\

\hline
\end{tabular}
\end{small}
\end{center}

~~

 {\bf Model 3.}
 The integer programming formulation of the minimum node weighted dominating set problem is:
 \[ \min \gamma^{w} (b) = \min \sum_{i=1}^{n} ~w(a_{i})  ~x_{i}\]
 \[s.t.~~~~ \forall a_{\zeta_{1}} \in (A \backslash B) ~~
 ~~\exists a_{\zeta_{2}} \in  B~ (i.e., ~ x_{a_{\zeta_{2}}} = 1)
 ~~such~~that ~~ (a_{\zeta_{1}},a_{\zeta_{2}}) \in R .\]

~~

 {\bf Model 4.}
 The integer programming formulation of the
 minimum vertex weighted connected dominating set problem is
 (here \(B_{c} \subseteq A \) is the connected dominating set):
 \[ \min \gamma^{w}_{c} (b) = \min \sum_{i=1}^{n} ~w(a_{i}) ~x_{i}\]
 \[s.t.~~~~ \forall a_{\zeta'}, a_{\zeta''} \in B_{c}
 ~~\exists ~~ path ~~ l(a_{\zeta'}, a_{\zeta''}) = <a_{\zeta'},..., a_{\zeta''}> , \]
 \[\forall a_{\zeta_{1}} \in (A \backslash B_{c}) ~~
 ~~\exists a_{\zeta_{2}} \in  B_{c}~ (i.e., ~ x_{a_{\zeta_{2}}} = 1)
 ~~such~~that ~~ (a_{\zeta_{1}},a_{\zeta_{2}}) \in R .\]

~~

\subsection{Basic multicriteria  models}

 Further,
 vertex weight vectors can be considered:
 ~\(\overline{w}(a_{i}) = (w(a^{1}_{i}),...,w^{\kappa}(a_{i}),...,w^{\mu} (a_{i}))\)
 ~(\( w^{\kappa}(a_{i})>0 ~ \forall \kappa = \overline{1,\mu}\),
 \(\forall a_{i} \in A\)).
 The following multicriteria (multiobjective) models can be examined
 (here the  Pareto-efficient solutions are searched for):

~~

 {\bf Model 5.} Multicritiria (multiobjective)
 minimum node weighted dominating set problem:
%
 \[
 \min \sum_{i=1}^{n} ~w^{1}(a_{i})~x_{i},...,
%
 \min \sum_{i=1}^{n} ~w^{\kappa}(a_{i})~x_{i},...,
%
 \min \sum_{i=1}^{n} ~w^{\mu}(a_{i})~x_{i},\]

 \[s.t.~~~~ \forall a_{\zeta_{1}} \in (A \backslash B) ~~
 ~~\exists a_{\zeta_{2}} \in  B~ (i.e., ~ x_{a_{\zeta_{2}}} = 1)
 ~~such~~that ~~ (a_{\zeta_{1}},a_{\zeta_{2}}) \in R .\]

~~

 {\bf Model 6.}
  Multicriteria (multiobjective) minimum node weighted connected dominating set problem:
%
%
 \[
  \min \sum_{i=1}^{n} ~w^{1}(a_{i}) ~x_{i},...,
 \min \sum_{i=1}^{n} ~w^{\kappa}(a_{i}) ~x_{i},...,
 \min \sum_{i=1}^{n} ~w^{\mu}(a_{i}) ~x_{i}, \]
 \[s.t.~~~~ \forall a_{\zeta'}, a_{\zeta''} \in B_{c}
 ~~\exists ~~ path ~~
 l (a_{\zeta'}, a_{\zeta''} ) = <a_{\zeta'},..., a_{\zeta''}> , \]
 \[\forall a_{\zeta_{1}} \in (A \backslash B_{c}) ~~
 ~~\exists a_{\zeta_{2}} \in  B_{c}~ (i.e., ~ x_{a_{\zeta_{2}}} = 1)
 ~~such~~that ~~ (a_{\zeta_{1}},a_{\zeta_{2}}) \in R .\]

~~

\subsection{Models of k-connected dominating set problems}

 In the case of \(k\)-connected dominating set the models are extended
 by an additional constraint for \(k\)-connectivity of dominating set:

~~

  {\bf Model 7.}
 The integer programming formulation of the basic
 minimum  \(k\) connected dominating set problem  is
 (here \(B_{c} \subseteq A \) is the \(k\)-connected dominating set):
 \[ \min \gamma_{c} (b) = \min \sum_{i=1}^{n} ~x_{i}\]
 \[s.t.~~~~ \forall a_{\zeta'}, a_{\zeta''} \in B_{c}
 ~~\exists ~~ k~~ vertex ~~disjoint ~~ paths\]
 \[ l^{1}( a_{\zeta'}, a_{\zeta''} ) = <a_{\zeta'},..., a_{\zeta''}>,...,
 ~ l^{k} (a_{\zeta'}, a_{\zeta''} ) = <a_{\zeta'},..., a_{\zeta''}> ; \]
 \[\forall a_{\zeta_{1}} \in (A \backslash B_{c}) ~~
 ~~\exists a_{\zeta_{2}} \in  B_{c}~ (i.e., ~ x_{a_{\zeta_{2}}} = 1)
 ~~such~~that ~~ (a_{\zeta_{1}},a_{\zeta_{2}}) \in R .\]

~~

 {\bf Model 8.}
 The integer programming formulation of the
 minimum vertex weighted \(k\)-connected connected dominating set problem is
 (here \(B_{c} \subseteq A \) is the \(k\)-connected dominating set):
 \[ \min \gamma^{w}_{c} (b) = \min \sum_{i=1}^{n} ~w(a_{i}) ~x_{i}\]
 \[s.t.~~~~ \forall a_{\zeta'}, a_{\zeta''} \in B_{c}
 ~~\exists ~~ k ~~  vertex ~~disjoint~~ paths\]
 \[ l^{1} (a_{\zeta'}, a_{\zeta''} ) = <a_{\zeta'},..., a_{\zeta''}>,...,
 ~ l^{k}(a_{\zeta'}, a_{\zeta''} ) = <a_{\zeta'},..., a_{\zeta''}> ; \]
 \[\forall a_{\zeta_{1}} \in (A \backslash B_{c}) ~~
 ~~\exists a_{\zeta_{2}} \in  B_{c}~ (i.e., ~ x_{a_{\zeta_{2}}} = 1)
 ~~such~~that ~~ (a_{\zeta_{1}},a_{\zeta_{2}}) \in R .\]
%


~~

  {\bf Model 9.}
  Multicriteria (multiobjective) minimum node weighted \(k\)-connected dominating set problem:
 \[
  \min \sum_{i=1}^{n} ~w^{1}(a_{i}) ~x_{i},...,
 \min \sum_{i=1}^{n} ~w^{\kappa}(a_{i}) ~x_{i},...,
 \min \sum_{i=1}^{n} ~w^{\mu}(a_{i}) ~x_{i}, \]
 \[s.t.~~~~ \forall a_{\zeta'}, a_{\zeta'} \in B_{c}
 ~~\exists ~~ k ~~  vertex ~~disjoint~~ paths\]
 \[ l^{1}(a_{\zeta'}, a_{\zeta''} ) = <a_{\zeta'},..., a_{\zeta'}>,...,
 ~ l^{k}(a_{\zeta'}, a_{\zeta''}) = <a_{\zeta'},..., a_{\zeta'}> ; \]
 \[\forall a_{\zeta_{1}} \in (A \backslash B_{c}) ~~
 ~~\exists a_{\zeta_{2}} \in  B_{c}~ (i.e., ~ x_{a_{\zeta_{2}}} = 1)
 ~~such~~that ~~ (a_{\zeta_{1}},a_{\zeta_{2}}) \in R .\]

~~

\subsection{Models of k-connected m-dominating set problems}

 In the case of \(k\)-connected \(m\) dominating set the models are extended
 by an additional constraints for \(k\)-connectivity and \(m\)-dominating:

~~~~

 {\bf Model 10.}
 The integer programming formulation of the basic
 minimum  \(k\)-connected \(m\)-dominating set problem  is
 (here \(B_{c} \subseteq A \) is the \(k\)-connected dominating set):
 \[ \min \gamma_{c} (b) = \min \sum_{i=1}^{n} ~x_{i}\]
 \[s.t.~~~~ \forall a_{\zeta'}, a_{\zeta''} \in B_{c}
 ~~\exists ~~ k ~~  vertex ~~disjoint~~ paths\]
 \[ l^{1}(a_{\zeta'}, a_{\zeta''} ) = <a_{\zeta'},..., a_{\zeta''}>,...,
 ~ l^{k}(a_{\zeta'}, a_{\zeta''} ) = <a_{\zeta'},..., a_{\zeta''}> ;\]
 \[\forall a_{\zeta_{1}} \in (A \backslash B_{c}) ~~
 ~~\exists
 ~m ~~vertices~~
 a_{\zeta_{1}}^{1}, ..., a_{\zeta_{1}}^{m} \in  B_{c}~
 (i.e., ~ x_{a_{\zeta_{1}}^{1}} = 1, ...,x_{a_{\zeta_{1}}^{m}} = 1
 )\]
 \[such~~that ~~ (a_{\zeta_{1}},a_{\zeta_{1}}^{1}) \in R,..., (a_{\zeta_{1}},a_{\zeta_{1}}^{m}) \in R  .\]

~~

 {\bf Model 11.}
 The integer programming formulation of the
 minimum vertex weighted \(k\)-connected connected
 \(m\)-dominating set problem is
 (here \(B_{c} \subseteq A \) is the \(k\)-connected dominating set):
 \[ \min \gamma^{w}_{c} (b) = \min \sum_{i=1}^{n} ~w(a_{i}) ~x_{i}\]
 \[s.t.~~~~ \forall a_{\zeta'}, a_{\zeta''} \in B_{c}
 ~~\exists ~~ k ~~  vertex ~~disjoint~~ paths\]
 \[ l^{1}(a_{\zeta'}, a_{\zeta''} ) = <a_{\zeta'},..., a_{\zeta''}>,...,
 ~ l^{k}(a_{\zeta'}, a_{\zeta''} ) = <a_{\zeta'},..., a_{\zeta''}> ;\]
 \[\forall a_{\zeta_{1}} \in (A \backslash B_{c}) ~~
 ~~\exists
 ~m ~~vertices~~
 a_{\zeta_{1}}^{1}, ..., a_{\zeta_{1}}^{m} \in  B_{c}~
 (i.e., ~ x_{a_{\zeta_{1}}^{1}} = 1, ...,x_{a_{\zeta_{1}}^{m}} = 1
 )\]
 \[such~~that ~~ (a_{\zeta_{1}},a_{\zeta_{1}}^{1}) \in R,..., (a_{\zeta_{1}},a_{\zeta_{1}}^{m}) \in R  .\]
%


~~

  {\bf Model 12.}
  Multicriteria (multiobjective) minimum node weighted \(k\)-connected
  \(m\)-dominating set problem:
 \[
  \min \sum_{i=1}^{n} ~w^{1}(a_{i}) ~x_{i},...,
 \min \sum_{i=1}^{n} ~w^{\kappa}(a_{i}) ~x_{i},...,
 \min \sum_{i=1}^{n} ~w^{\mu}(a_{i}) ~x_{i}, \]
 \[s.t.~~~~ \forall a_{\zeta'}, a_{\zeta''} \in B_{c}
 ~~\exists ~~ k ~~  vertex ~~disjoint~~ paths\]
 \[ l^{1}(a_{\zeta'}, a_{\zeta''} ) = <a_{\zeta'},..., a_{\zeta''}>,...,
 ~ l^{k}(a_{\zeta'}, a_{\zeta''} ) = <a_{\zeta'},..., a_{\zeta''}> ;\]
 \[\forall a_{\zeta_{1}} \in (A \backslash B_{c}) ~~
 ~~\exists
 ~m ~~vertices~~
 a_{\zeta_{1}}^{1}, ..., a_{\zeta_{1}}^{m} \in  B_{c}~
 (i.e., ~ x_{a_{\zeta_{1}}^{1}} = 1, ...,x_{a_{\zeta_{1}}^{m}} = 1
 )\]
 \[such~~that ~~ (a_{\zeta_{1}},a_{\zeta_{1}}^{1}) \in R,..., (a_{\zeta_{1}},a_{\zeta_{1}}^{m}) \in R  .\]

~~

\subsection{Problems with multiset estimates}

\subsubsection{Interval multiset estimates}

 The fundamentals of the multiset theory
 are  presented in \cite{knuth98,yager86}.
  Here a brief description of interval multiset estimates
  from \cite{lev15} is described.
 The approach consists in assignment of elements (\(1,2,3,...\))
 into an ordinal scale \([1,2,...,l]\).
 In the obtained  multi-set based estimate,
  a basis set involves all levels of the ordinal scale:
 \(\Omega = \{ 1,2,...,l\}\) (the levels are linear ordered:
 \(1 \succ 2 \succ 3 \succ ...\)) and
 the assessment problem (for each alternative/object under assessment)
 consists in selection of a multiset over set \(\Omega\)
 while taking into account two conditions:

 {\it 1.} cardinality of the selected multiset equals a specified
 number of elements \( \eta = 1,2,3,...\)
 (i.e., multisets of cardinality \(\eta \) are considered);

 {\it 2.} ``configuration'' of the multiset is the following:
 the selected elements of \(\Omega\) cover an interval over scale \([1,l]\)
 (i.e., ``interval multiset estimate'').

 An estimate \(e\)  for an alternative/object \(a\)
 (e.g., \(a \in A\) )
  is (scale \([1,l]\), position-based form or position form):
 \(e(a) = (\eta_{1},...,\eta_{\iota},...,\eta_{l})\),
 where \(\eta_{\iota}\) corresponds to the number of elements at the
 level \(\iota\) (\(\iota = \overline{1,l}\)), or
 \(e(a) = \{ \overbrace{1,...,1}^{\eta_{1}},\overbrace{2,...2}^{\eta_{2}},
 \overbrace{3,...,3}^{\eta_{3}},...,\overbrace{l,...,l}^{\eta_{l}}
 \}\).
 The number of multisets of cardinality \(\eta\),
 with elements taken from a finite set of cardinality \(l\),
 is called the
 ``multiset coefficient'' or ``multiset number''
  (\cite{knuth98,yager86}):
 ~~\( \mu^{l,\eta} =
   \frac{l(l+1)(l+2)... (l+\eta-1) } {\eta!}  \).
 This number corresponds to possible estimates
 (without taking into account interval condition 2).
 In the case of condition 2, the number of estimates is decreased.
 The certain assessment problem based on interval multiset estimates
 can be denoted as the following: ~\(P^{l,\eta}\).

  In addition, basic operations over multiset estimates are used:
 integration, vector-like proximity, aggregation, and alignment
 \cite{lev15}.
%
 Integration of estimates (mainly, for composite systems)
 is based on summarization of the estimates by components (i.e.,
 positions).
  Let us consider \(n\) estimates (position form):~
 estimate \(e^{1} = (\eta^{1}_{1},...,\eta^{1}_{\iota},...,\eta^{1}_{l})
 \),
  {\bf ...},
 estimate \(e^{\kappa} = (\eta^{\kappa}_{1},...,\eta^{\kappa}_{\iota},...,\eta^{\kappa}_{l})
 \),
  {\bf ...},
 estimate \(e^{n} = (\eta^{n}_{1},...,\eta^{n}_{\iota},...,\eta^{n}_{l})
 \).
 The integrated estimate is:~
 estimate \(e^{I} = (\eta^{I}_{1},...,\eta^{I}_{\iota},...,\eta^{I}_{l})
 \),
 where
 \(\eta^{I}_{\iota} = \sum_{\kappa=1}^{n} \eta^{\kappa}_{\iota} ~~ \forall
 \iota = \overline{1,l}\).
 In fact, the operation \(\biguplus\) is used for multiset
 estimates:~
 \(e^{I} = e^{1} \biguplus ... \biguplus e^{\kappa} \biguplus ... \biguplus
 e^{n}\).

 The vector-like proximity of two multiset based estimates
 ~\(e(a_{1})\), \(e(a_{2})\) ~(e.g., ~\(a_{1},a_{2} \in A\)) is:
 \[\delta ( e(A_{1}), e(A_{2})) = (\delta^{-}(A_{1},A_{2}),\delta^{+}(A_{1},A_{2})),\]
 where vector components are:
 (i) \(\delta^{-}\) is the number of one-step changes:
 element of quality \(\iota + 1\) into element of quality \(\iota\) (\(\iota = \overline{1,l-1}\))
 (this corresponds to ``improvement'');
 (ii) \(\delta^{+}\) is the number of one-step changes:
 element of quality \(\iota\) into element of quality  \(\iota+1\) (\(\iota = \overline{1,l-1}\))
 (this corresponds to ``degradation'').
 It is assumed:
 ~\( | \delta ( e(A_{1}), e(A_{2})) | = | \delta^{-}(A_{1},A_{2}) | + |\delta^{+}(A_{1},A_{2})|
 \).
%

 For aggregation (as an analogue of summarization)
 of the specified set of multiset based estimates
 \(E = \{ e_{1},...,e_{\kappa},...,e_{n}\}\),
 two types of medians are considered.
  Here \(\overline{E} \) corresponds to the
  the set of all possible multiset based estimates
 (\( E  \subseteq \overline{E} \)).
%
 The median estimates are:

  (a) ``generalized median'':
 ~~\(M^{g} =   \arg \min_{M \in \overline{E}}~
   \sum_{\kappa=1}^{n} ~  | \delta (M, e_{\kappa}) |; ~~\)

 (b) ``set median'':
 \( M^{s} =   \arg \min_{M\in E} ~
   \sum_{\kappa=1}^{n} ~ | \delta (M, e_{\kappa}) |\).

\subsubsection{Models with multiset estimates}

 Here vertex weight vectors
 ~\(\overline{w}(a_{i}) = (w(a^{1}_{i}),...,w^{\kappa}(a_{i}),...,w^{\mu} (a_{i}))\)
 ~(\( w^{\kappa}(a_{i})>0 ~\forall \kappa = \overline{1,\mu}\),
 \(\forall a_{i} \in A\))
 is transformed (compressed) into multiset based estimate
  ~\(e(a_{i})\).
 The following models can be considered:

~~

 {\bf Model 13.}
 Minimum node weighted dominating set problem with multiset estimates
 (the objective function is based on ``generalized median''):
%
 \[ \min~  M^{g} =   \arg \min_{M \in \overline{E}}~
   \sum_{\kappa=1}^{n} ~  | \delta (M, e_{\kappa}) |; ~~\]
 \[s.t.~~~~ \forall a_{\zeta_{1}} \in (A \backslash B) ~~
 ~~\exists a_{\zeta_{2}} \in  B~ (i.e., ~ x_{a_{\zeta_{2}}} = 1)
 ~~such~~that ~~ (a_{\zeta_{1}},a_{\zeta_{2}}) \in R .\]

~~

  {\bf Model 14.}
 Minimum node weighted connected dominating set problem with multiset estimates
 (the objective function is based on ``generalized median''):
 \[ \min~   M^{g} =   \arg \min_{M \in \overline{E}}~
   \sum_{\kappa=1}^{n} ~  | \delta (M, e_{\kappa}) |; ~~\]
 \[s.t.~~~~ \forall a_{\zeta'}, a_{\zeta''} \in B_{c}
 ~~\exists ~~ path ~~
 l(a_{\zeta'}, a_{\zeta''}) = <a_{\zeta'},..., a_{\zeta''}> , \]
 \[\forall a_{\zeta_{1}} \in (A \backslash B_{c}) ~~
 ~~\exists a_{\zeta_{2}} \in  B_{c}~ (i.e., ~ x_{a_{\zeta_{2}}} = 1)
 ~~such~~that ~~ (a_{\zeta_{1}},a_{\zeta_{2}}) \in R .\]

~~

  {\bf Model 15.}
  Multicriteria (multiobjective) minimum node weighted \(k\)-connected dominating set problem
  with multiset estimates
  (the objective function is based on ``generalized median''):
 \[ \min~   M^{g} =   \arg \min_{M \in \overline{E}}~
   \sum_{\kappa=1}^{n} ~  | \delta (M, e_{\kappa}) |; ~~\]
 \[s.t.~~~~ \forall a_{\zeta'}, a_{\zeta'} \in B_{c}
 ~~\exists ~~ k ~~  vertex ~~disjoint~~ paths\]
 \[ l^{1}(a_{\zeta'}, a_{\zeta''} ) = <a_{\zeta'},..., a_{\zeta'}>,...,
 ~ l^{k}(a_{\zeta'}, a_{\zeta''}) = <a_{\zeta'},..., a_{\zeta'}> ; \]
 \[\forall a_{\zeta_{1}} \in (A \backslash B_{c}) ~~
 ~~\exists a_{\zeta_{2}} \in  B_{c}~ (i.e., ~ x_{a_{\zeta_{2}}} = 1)
 ~~such~~that ~~ (a_{\zeta_{1}},a_{\zeta_{2}}) \in R .\]

~~

  {\bf Model 16.}
  Multicriteria (multiobjective) minimum node weighted \(k\)-connected
  \(m\)-dominating set problem with multiset estimates
  (the objective function is based on ``generalized median''):
%
 \[ \min~   M^{g} =   \arg \min_{M \in \overline{E}}~
   \sum_{\kappa=1}^{n} ~  | \delta (M, e_{\kappa}) |; ~~\]
 \[s.t.~~~~ \forall a_{\zeta'}, a_{\zeta''} \in B_{c}
 ~~\exists ~~ k ~~  vertex ~~disjoint~~ paths\]
 \[ l^{1}(a_{\zeta'}, a_{\zeta''} ) = <a_{\zeta'},..., a_{\zeta''}>,...,
 ~ l^{k}(a_{\zeta'}, a_{\zeta''} ) = <a_{\zeta'},..., a_{\zeta''}> ;\]
 \[\forall a_{\zeta_{1}} \in (A \backslash B_{c}) ~~
 ~~\exists
 ~m ~~vertices~~
 a_{\zeta_{1}}^{1}, ..., a_{\zeta_{1}}^{m} \in  B_{c}~
 (i.e., ~ x_{a_{\zeta_{1}}^{1}} = 1, ...,x_{a_{\zeta_{1}}^{m}} = 1
 )\]
 \[such~~that ~~ (a_{\zeta_{1}},a_{\zeta_{1}}^{1}) \in R,..., (a_{\zeta_{1}},a_{\zeta_{1}}^{m}) \in R  .\]
%


\section{Conclusion}

 In the paper,
 a brief survey on dominating set problems is presented.
 The integer programming formulations
 of dominating set problems are described.
 Some new multicriteria models and models based on multiset estimates
 are considered as well.

 It may be reasonable to point out
 the following future research directions:
 (1) consideration of other application domains for the
  dominating set problems;
 (2) examination of multistage solving strategies;
 (3) further examination
 of various \(k\)-connected \(m\)-dominating set problems;
 (4) additional study of the described combinatorial models
 and their various versions
 (including studies of domination structures as trees, etc.);
 (5) special future studies of multi-hop dominating set problems
 on various graphs (including applications in networking);
 (6) design of a special software package
 (i.e., decision support system)
  for the dominating set problems
  (including descriptions of problems,
 descriptions of applied domains,
   solving methods/procedures, examples);
 and
  (7) using the
  described optimization models for
 dominating set problems in education
 (e.g., student projects).


  The author states that there is no conflict of interest.



~~

 Author work address:~
 Mark Sh. Levin, Inst. for Information Transmission Problem,
 Russian Academy of Sciences,
 19 Bolshoy Karetny lane, Moscow 127051, Russia

~~

 http://www.mslevin.iitp.ru/ ~~~ email: mslevin@acm.org

~~

 Author home address:~
 Mark Sh. Levin, Sumskoy Proezd 5-1-103, Moscow 117208, Russia

\end{document}